\documentclass[amsmath,amssymb,12pt,aps]{revtex4-2}
\usepackage{amsfonts}
\usepackage[utf8]{inputenc} 
\usepackage{graphicx}
\usepackage{float}
\usepackage{natbib}
\usepackage{xcolor}
\usepackage{dcolumn}
\usepackage[english]{babel}
\newcommand{\be}{\begin{equation}}
\newcommand{\ee}{\end{equation}}

\begin{document}

\title{Synchronization of identical oscillators on a sphere: exact results with external forces and higher-order interactions}

\author{Guilherme S. Costa$^1$}
\author{Marcel Novaes$^{2,4}$}	
\author{Ricardo Fariello$^3$}
\author{Marcus A.~M.~de Aguiar$^{1,4}$}

\affiliation{$^1$ ICTP South American Institute for Fundamental Research \& Instituto de Física Teórica - UNESP, 01140-070, São Paulo, Brazil }
\affiliation{$^2$Instituto de F\'isica, Universidade Federal de Uberl\^andia, 38408-100, Uberl\^andia, MG, Brazil}
\affiliation{$^3$ Departamento de Ci\^encias da Computa\c{c}\~ao, Universidade Estadual de Montes Claros, 39401-089, Montes Claros, MG, Brazil} 
\affiliation{$^4$Instituto de F\'isica Gleb Wataghin, Universidade Estadual de Campinas, 13083-970, Campinas, SP, Brazil}

\begin{abstract}

We study the dynamics of the Kuramoto model on the sphere under higher-order interactions and an external periodic force. For identical oscillators, we introduce a novel way to incorporate three- and four-body interactions into the dynamics of the order parameter, allowing for a full dimensional reduction of this system. We discuss how such reduction can be implemented in two different ways and how they are related. When restricted to the equator, the dynamics is similar to that of the usual Kuramoto model, up to an interesting renormalization of the coupling constants. Outside this plane, the motion reduces to a two-parameter set of periodic orbits. We also locate the bifurcation curves of the system as functions of different parameters. 

\end{abstract}

\maketitle

\section{Introduction}

The study of synchronization in large populations of oscillatory units is a topic of interest in several areas of
science. It appears in the collective behavior of animals, from fireflies \cite{Moiseff181} to humans \cite{hadar1985head,neda2000sound}, in the firing of neurons in the
brain \cite{Mackay1997,cumin2007generalising,Deco2011,Schmidt2015}, pacemaker heart cells \cite{Michaels704} and in a variety of artificial oscillators, such as pendulum clocks \cite{oliveira2014huygens} and metronomes \cite{Pantaleone2002}. Inspired
by Winfree's intuition that weakly coupled oscillatory systems could be described only by their phases
\cite{winfree1967biological}, Kuramoto developed a model that would become a paradigm in the study of synchronization
\cite{Kuramoto1975,Kuramoto1984}.  

In the Kuramoto model the oscillators interact with each other according to
\be \dot{\theta}_i = \omega_i + \frac{K}{N} \sum_{j=1}^N \sin{(\theta_j-\theta_i)},\ee 
where $\theta_i$ is the phase of the $i$-th oscillator and $K$ is a coupling constant. The natural frequencies $\omega_i$ are usually drawn from a symmetric unimodal distribution $g(\omega)$ centered on $\omega_0$. Kuramoto showed that a phase transition exists in the large-$N$ limit, with oscillators synchronizing at frequency $\omega_0$ as $K$ becomes larger than some critical value $K_c$. The transition is continuous (second order) for smooth distributions $g(\omega)$ and can be characterized by the complex number
\begin{equation}
	z = r e^{i \psi} \equiv \frac{1}{N} \sum_{j=1}^N e^{i\theta_j}.
	\label{paraord}
\end{equation}
The order parameter $r=Re(z)$ goes from zero, when the motion is incoherent, to one, for full synchronization. 

Oscillatory systems are often subjected to external stimuli that can significantly alter their behavior. Flashing lights, for example, can trigger epileptic seizures \cite{harding1994photosensitive} and artificial pacemakers can control synchronization of heart cells \cite{Reece2012}. The external periodic forces that act on the oscillatory units introduce a tendency to synchronize at the external frequency and compete with the internal dynamics \cite{Sakaguchi1988,Childs2008,Hindes2015,Lizarraga2020,PhysRevLett.131.030401,odor2023synchronization}. Childs and Strogatz \cite{Childs2008} found that the collective states of oscillators modeled by the Kuramoto equations and subjected to a periodic external force can undergo several phase transitions as the intensity and frequency of the force are varied. Such detailed analytical results were derived in the limit of infinitely many oscillators, in which the ansatz proposed by Ott and Antonsen \cite{Ott2008} can be applied. This framework consists in restricting the distribution of oscillators to a specific functional form that leads to equations of motion for the order parameter, drastically reducing the complexity of the system. 

In the past 30 years, the Kuramoto model has been explored and generalized in many ways \cite{Rodrigues2016}. Particularly, its extension to higher dimensions is of extreme importance for this work. The idea, first put forward by Lohe \cite{lohe2009non}, can be understood as follows: the phase $\theta_i$ of an oscillator in the Kuramoto model can be viewed as a particle moving in the unit circle. The position of the particle corresponds to the phase of oscillation. Lohe showed that the Kuramoto equations can be naturally extended to allow the particles to move on the surface of unit spheres in any number of dimensions. In 2D the sphere is the unit circle and the Kuramoto model is recovered. In 3D, on the other hand, particles would be described by a polar and an azimuthal angle, instead of a single phase as in the Kuramoto model. Angles can represent, for instance, the flight direction of birds or insects in a swarm. One of the most striking features of the 3D Kuramoto model is the appearance of a discontinuous phase transition to synchronization at $K=0$ \cite{2019continuous}, contrasting with the continuous transition at $K=K_c > 0$ of the original model. Complexity reduction proposals similar to the Ott-Antonsen ansatz were also investigated in
Refs.~\cite{Tanaka2014,chandra2019complexity,lipton2021Kuramoto,barioni2021complexity,barioni2021ott}. 

Despite advances in understanding synchronization within the framework of the Kuramoto model, recent studies have pointed out the importance of many-body interactions in many dynamical processes, as opposed to the usual pairwise (two-body) functional form $\sin (\theta_j -\theta_i)$.  \cite{tanaka2011multistable,bick2016chaos,battiston2020networks,dutta2023impact,leon2024higher}. Interactions involving the collective action of three or more agents can be found in different areas, such as neuroscience \cite{ganmor2011sparse,petri2014homological,giusti2015clique,reimann2017cliques,sizemore2018cliques}, ecology \cite{grilli2017higher,ghosh2024chimeric}, biology \cite{sanchez2019high}, and social sciences \cite{benson2016higher,de2020social}. Pairwise interactions correspond to the notion of a graph where oscillators are nodes and edges represent the interaction. Higher-order interactions, on the other hand, correspond to a simplicial complex or hypergraph that contain, besides nodes and edges, triangles, tetrahedra, etc. For example, in the Kuramoto model, one can add to the equation of motion of $\theta_i$ interaction terms proportional to $K_2\sin(2\theta_j-\theta_k-\theta_i)$. It has been shown that a large enough $K_2$ can lead to bi-stability and the appearance of a first-order phase transition.

In a recent paper, we investigated the joint effects of asymmetric higher-order interactions and external periodic forces on the Kuramoto model. In this article, our aim is to extend these results to 3D. There are two main difficulties with such a program. The first is that the time dependence of a periodic external force cannot be eliminated by changing to a rotating reference, as 3D matrices generally do not commute. However, analytical results can still be obtained if the external force and natural frequencies are restricted to point in the same direction. The second issue refers to the way three-body interactions are represented in the sphere \cite{fariello2024third}. In the limit $N \rightarrow \infty$, these interactions cannot be written just in terms of the system order parameter, as they depend on a correlation matrix. We get around this problem by proposing an approximation that allows us to derive analytical results. The approximation is justified by numerical results. For the case of identical oscillators, our main results are: (i) when the motion is restricted to a plane, the system behaves similarly to the 2D model, although with re-scaled parameters; and (ii) there is a two-parameter family of periodic solutions that is not restricted to two dimensions even if force and natural frequencies are in the same direction. We also construct the complete bifurcation diagram of the system. 

This paper is organized as follows. In Section II we discuss how asymmetric higher-order terms can be included in the spherical Kuramoto model. In Section III we introduce the analogue of the Ott-Antonsen ansatz in 3D and a further ansatz to treat three-body interactions that allows us to derive analytical equations of motion for the order parameter of this model. In Section IV, an external force is added to the system, enriching its dynamics. Section V is dedicated to some comparisons between the planar and spherical models, together with some simple solutions of the model. Section VI discusses bifurcation scenarios.

\section{Asymmetric Higher order interactions in 3D}
\label{secprelim}

\subsection{Traditional Kuramoto model}
\label{secprlima}

The problem of Kuramoto oscillators with asymmetric higher-order interactions was treated in detail in \cite{skardal2020higher}. The dynamics adhere to the equations
\begin{eqnarray}
	\dot{\theta}_i &=& \omega_i + \frac{K_1}{N} \sum_{j=1}^N \sin{(\theta_j-\theta_i)} + 
	\frac{K_2}{N^2} \sum_{j,k=1}^N \sin{(2\theta_j-\theta_k -\theta_i)} \nonumber \\
	&+& \frac{K_3}{N^3} \sum_{j,k,m=1}^N \sin{(\theta_j-\theta_k +\theta_m-\theta_i)} ,
	\label{kuramoto}
\end{eqnarray}
where $K_1$, $K_2$ and $K_3$ are the coupling constants that control the relative intensity of $1$-, $2$- and $3$-simplexes, respectively. Recalling the definition of the order parameter, Eq.~(\ref{paraord}), and introducing
\begin{equation}
	z_2 = \frac{1}{N} \sum_j e^{2 i \theta_j} \equiv r_2 e^{i \psi_2},
	\label{z2def}
\end{equation}
Eq.~(\ref{kuramoto}) can be written as
\begin{equation}
	\dot{\theta}_i =\omega_i + K_1 r \sin(\psi - \theta_i) +  K_2 r r_2 \sin{(\psi_2-\theta_i-\psi)} + K_3 r^3 \sin(\psi-\theta_i).
	\label{kuramoto25}
\end{equation}

The main advantage of considering asymmetric higher-order interactions is that they allow direct application of the Ott-Antonsen ansatz \cite{Ott2008} and, therefore, analytical treatment of the system in the thermodynamic limit. Assuming a Lorentzian distribution of natural frequencies with width $\Delta$, Skardal and Arenas \cite{skardal2020higher} derived an exact equation for the modulus of the order parameter:
\begin{equation}\label{rdot2D}
	\dot{r} = -\Delta r + \frac{r}{2}(1-r^2) (K_1 + K_{23} r^2),
\end{equation}
where $K_{23}=K_2+K_3$. A complete analysis of the bifurcation diagram as a function of the parameters $K_1$ and $K_{23}$
was also provided, showing that if the $K_{23}$ is large enough, it can lead to bi-stability, promoted by a saddle-node bifurcation, and the appearance of a first-order phase transition. 

\subsection{Vector formulation}
\label{secprlimb}

Equation~(\ref{kuramoto25}) can be formulated in terms of unit vectors $\vec{\sigma_i} = (\cos{\theta_i},\sin{\theta_i}) \equiv (\sigma_{ix},\sigma_{iy})$ and then extended to higher dimensions \cite{2019continuous,barioni2021complexity,fariello2024third}. Defining
\be \vec{r} \equiv \vec{R}_1= (r\cos\psi, r\sin\psi),\ee 
\be \vec{R}_2 =  \frac{2}{N} \sum_j (\vec{r} \cdot \vec{\sigma}_j) \vec{\sigma}_j  - \vec{r}\ee
and
\be\vec{R}_3 = r^2 \vec{r},\ee 
Eq.~(\ref{kuramoto25}) can be written in vector form as \cite{fariello2024third}
\begin{equation}
	\frac{d \vec{\sigma_i}}{d t} = \mathbf{W}_i \vec{\sigma_i} + K_1
     [\vec{r} -(\vec{\sigma}_i \cdot \vec{r} ) \vec{\sigma}_i] +
    \sum_{j=2}^3 K_j [\vec{R}_j -(\vec{\sigma}_i \cdot \vec{R}_j) \vec{\sigma}_i],
    \label{kuracomplete1}
\end{equation}
where
\begin{equation}
	W_i = \left( 
	\begin{array}{cc}
		0 & -\omega_i \\
		\omega_i & 0
	\end{array}
	\right).
	\label{wmat}
\end{equation}

Equation~(\ref{kuracomplete1}) can now be extended to higher dimensions by simply considering unit vectors $\vec\sigma_i$ in $D$ dimensions, rotating on the surface of the corresponding $(D-1)$ unit sphere \cite{2019continuous,Strogatz2019higher}. The particles are represented by $D-1$ spherical angles, generalizing the single phase $\theta_i$ of the original model. The matrices $\mathbf{W}_i$ become $D \times D$ antisymmetric matrices containing the $D(D-1)/2$ natural frequencies of each oscillator. 

For the particular case of 3 dimensions, we may write
\begin{equation}
	\mathbf{W}_i = \left( 
	\begin{array}{ccc}
		0 & \omega_{i3}  &  -\omega_{i2} \\
		-\omega_{i3} & 0  &  \omega_{i1} \\
		\omega_{i2} & -\omega_{i1} & 0 
	\end{array}
	\right)
	\label{wmat3a}
\end{equation}
and $ \mathbf{W}_i \vec\sigma_i = \vec \omega_i \times \vec \sigma_i$ where $\vec{\omega}_i = (\omega_{i1},\omega_{i2},\omega_{i3})$. Likewise, the distribution of natural frequency matrices $G(\mathbf{W})$ can be written as a distribution of vectors $G(\vec{\omega})$. Eq.~(\ref{kuracomplete1}) then becomes
\begin{equation}
	\frac{d \vec{\sigma_i}}{d t} = \vec \omega_i \times \vec \sigma_i + K_1
     [\vec{r} -(\vec{\sigma}_i \cdot \vec{r}) \vec{\sigma}_i] +
    \sum_{j=2}^3 K_j [\vec{R}_j -(\vec{\sigma}_i \cdot \vec{R}_j) \vec{\sigma}_i],
    \label{kuracomplete3d}
\end{equation}
where $\vec\sigma_i = (\sigma_{ix},\sigma_{iy},\sigma_{iz})$.

The 3 and 4-dimensional cases without external forces were recently studied in \cite{fariello2024third}, where the coupling constants were generalized to coupling matrices \cite{buzanello2022matrix,de2023generalized}. When $K_2=K_3=0$ the order parameter has a discontinuous (first-order) transition at $K_1=0$, with $r(K_1=0^-)=0$ and $r(K_1=0^+)=0.5$ \cite{2019continuous}. Because $\vec R_3 = r^2 \vec{r}$, four-body interactions have a simple contribution to $r$, since it amounts to making $K_1 \rightarrow K_1 + r^2 K_3$. For $K_1 <0$, $r=0$ and $K_3$ do not contribute and as $r \rightarrow 1$ it makes $K_1 \rightarrow K_1 + K_3$, facilitating synchronization.

The role of 3-body interactions is more interesting. Consider $D=3$. Following \cite{fariello2024third}, we rewrite $\vec{R}_2 =  (2 \mathbf{U} - \mathbf{1})  \vec{r}$ where $\mathbf{U}=\frac{1}{N} \sum_j ( \vec{\sigma}_j \vec{\sigma}_j^T)$ and $\mathbf{1}$ is the $3 \times 3$ identity matrix. When $r=0$ the vectors $\vec\sigma_j$ are uncorrelated and all off-diagonal matrix elements of $\mathbf{U}$ are zero. The diagonal terms, on the other hand, must add up to one. Therefore, when $r=0$ we find $\mathbf{U}= \frac{1}{3} \mathbf{1}$ and $\vec{R}_2 = -\frac{1}{3} \vec{r}$. When $r \rightarrow 1$, on the other hand, $\mathbf{U} \rightarrow \mathbf{1}$ and $\vec{R}_2 =\vec{r}$.

The qualitative conclusion of this discussion for $D=3$ is that, when $r \approx 0$, the effect of the three- and four-body interaction terms is to make $K_1 \rightarrow K_1 - K_2/3$, making synchronization harder and requiring $K_1 >  K_2/3$. For $r\approx 1$, on the other hand, the effect is additive and $K_1 \rightarrow K_1 + K_2 + K_3$, facilitating synchronization. We will use this analysis later.


%
%

\section{Dimensional reduction on the sphere}
\label{secoa}

\subsection{Traditional Ott-Antonsen ansatz}

In the $N\to\infty$ limit, the system can be described by a function $f(\theta,\omega,t)$ that represents the density of oscillators with natural frequency $\omega$ that are at position $\theta$ at time $t$. It can be written as a Fourier series
\be f=\frac{G(\omega)}{2\pi}\left(1+\sum_{n>0} (f_ne^{in\theta}+f_n^*e^{-in\theta}) \right),\ee
where $G(\omega)$ is the distribution of natural frequencies. 

The Ott-Antonsen ansatz consists in choosing the family of densities for which $f_n=\alpha^n$ for some $\alpha(\omega,t)$. This leads to 
\be 
f=\frac{G(\omega)}{2\pi}\frac{1-|\alpha|^2}{|1-\alpha^*e^{i\theta}|^2}
\label{fourier}
\ee 
and to an equation of motion for the parameter $\alpha$. 

This ansatz therefore promotes a dimensional reduction of the system in such a way that the $N$ coupled equations of motion are ultimately described by a pair of coupled equations.

When all oscillators are equal, it can be shown that this equation acts as the generator of hyperbolic M\"obius transformations, revealing the group theory framework behind the ansatz, which in this case becomes exact.

\subsection{Two ansatzes on the sphere}

In the sphere, consider $f(\vec\omega,\theta,\phi,t)$ as the density of oscillators with natural frequency $\vec{\omega}$ at latitude and longitude $\theta$, $\phi$. There have been two suggestions of dimensional reductions that are analogous to the Ott-Antonsen ansatz, both of which use a vector parametrization \be \vec\rho(\vec\omega,t) = \rho (\sin\Theta\cos\Phi,\sin\Theta\sin\Phi, \cos\Theta)\ee instead of the previous complex number $\alpha$.

In both cases, the density becomes perfectly localized for $\rho =1$, indicating full synchrony of the oscillators with the same natural frequency, while $\rho =0$ corresponds to oscillators uniformly spread over the sphere. The time evolution of the density is then mapped into a dynamical equation for the vector parameter $\vec{\rho}$.

The first one, developed in \cite{chandra2019complexity,lipton2021Kuramoto}, is based on the corresponding theory of hyperbolic M\"obius transformations. In this approach, the density is taken as proportional to the hyperbolic Poisson kernel
\be
\label{ansatz1} 
f_P(\vec\omega,\theta,\phi,t) =\frac{G(\vec\omega)}{4\pi}\frac{(1-\rho^2)^2}{|\vec{\rho}-\hat{\sigma}|^4},
\ee
where $\hat{\sigma} = (\sin\theta\cos\phi,\sin\theta\sin\phi, \cos\theta)$.

The second one, developed in \cite{barioni2021complexity,barioni2021ott}, replaces the Fourier series (\ref{fourier}) by a spherical harmonics expansion,
\be
f_Y(\vec\omega,\theta,\phi,t) =\frac{G(\vec\omega)}{4\pi}\left[1+4\pi\sum_{l=1}^\infty \sum_{m=-l}^l f_{lm} Y_{lm}(\theta,\phi)\right]
\ee
and then chooses the coefficients as $f_{lm}=\rho^l \, Y^*_{lm}(\Theta,\Phi)$, which leads to a different density,
\begin{equation}
		f_Y(\vec\omega,\theta,\phi,t) = \frac{G(\vec\omega)}{4\pi}\frac{(1-\rho^2)}{|\vec{\rho}-\hat{\sigma}|^3}.
		\label{anssol}
\end{equation}

\subsection{Reduced equations}
\label{secoa1}

Imposing that these densities satisfy the continuity equation requires that
\begin{equation}
	\frac{\partial f}{\partial t} + \frac{1}{\sin\theta} \frac{\partial}{\partial \theta}
	(f\sin\theta v_\theta)+ \frac{1}{\sin\theta}\frac{\partial}{\partial\phi}(f v_\phi) = 0
    \label{oacont}
\end{equation}
for the velocity field 
\be 
\vec{v} = \vec{\omega} \times \vec{\sigma}+ \vec q -(\vec\sigma \cdot \vec q)\vec\sigma.
\label{vfield}
\ee
In both cases, this leads to
\begin{equation}
		\dot{\vec{\rho}} = \vec{\omega} \times \vec{\rho}  + \frac{1}{2} (1+\rho^2) \vec{q}  -  (\vec{\rho} \cdot \vec{q} ) \vec{\rho}
	\label{eqm1fa}
\end{equation}   
where $\vec{q} = K \vec{r}$ for the Kuramoto model with pair interactions. However, $f_Y$ further requires that the interaction vector $\vec{q}$ be displaced in the direction of $\hat{\sigma}$ by $\vec{q} \rightarrow \vec{q} + (K/4)[(\hat{\rho}-\hat{\sigma})\cdot \vec{r}] \hat{\sigma}$ \cite{barioni2021ott}. This does not change the velocity field, as terms in the radial direction are automatically canceled by the form of interaction being tangential to the sphere.

The relation between $\vec{r}$ and the vector $\vec{\rho}$ is different in these approaches. For $f_P$, it is given by
\be\label{r1}
\vec{r} = \int \vec{\rho}\, h(\rho) g(\vec\omega) d^3 \omega
\ee
where $\rho = |\vec{\rho}|$ and
\be 
h(\rho)=\frac{1+\rho^2}{2\rho^2}+\frac{(1-\rho^2)^2}{4\rho^3}\log\left(\frac{1-\rho}{1+\rho}\right).
\ee
For $f_Y$, on the other hand, the relation is simpler:
\be\label{r2}
\vec{r} = \int \vec{\rho}\, g(\vec\omega) d^3 \omega.
\ee
The two distributions coincide if the initial conditions are chosen such that $\rho(0)=1$. It is easy to see that this implies $\rho(t)=1$ for all $t$ and hence $h(\rho)=1$. This choice corresponds to starting all oscillators with the same natural frequency in full synchrony. Simulations with the Kuramoto model show that even if $\rho(0) < 1$ they converge to 1 in equilibrium. In this case, both $f_P$ and $f_Y$ tend to $\delta(\hat{\sigma}(\theta,\phi)-\hat{\rho}(\Theta,\Phi))$. The displacement of $\vec{q}$ required for $f_Y$ is irrelevant in this case, since it vanishes for $\hat{\sigma} = \hat{\rho}$ and the distribution is zero for $\hat{\sigma} \neq \hat{\rho}$.

However, it is not clear that the ansatzes coincide when higher-order terms and external forces are added to the system, which is of interest here. Then it is possible to have partial synchronization even for identical oscillators and condition $|\vec{\rho}|=1$ is not generally true. For the particular case of identical oscillators, which we shall consider in this paper, the use of Eq.~(\ref{ansatz1}) and $\vec{r}=h(\rho)\vec{\rho}$ results in a complicated equation for the order parameter. Our strategy, therefore, is to use the ansatz proposed in \cite{barioni2021ott} and $\vec{r}=\vec{\rho}$ to derive simple analytical results and compare them with numerical solutions of the Kuramoto equations and with numerical evaluations of the ansatz (\ref{ansatz1}). Since $|h(\rho)-1|$ is zero for $\rho=0$ and $\rho=1$ and is small for $0 < \rho < 1$, we expect the differences to be small. 

To further investigate the relationship between ansatzes, we also considered a very narrow distribution $g(\vec{\omega})$ and integrated Eqs.~(\ref{eqm1fa}) and (\ref{r2}), discretizing the integral by sampling several values of $\vec{\omega}$ according to $g(\vec{\omega})$ and setting initial conditions for the corresponding vectors $\vec{\rho}(\omega,t)$ such that $|\vec{\rho}(\omega,0)|=1$. The resulting order parameter turned out to be independent of the ansatz used. Notice that, when using $f_Y$, averaging was necessary even for identical oscillators in order to obtain an accurate result. However, we shall see that the analytical results achieved without the average were simple and insightful.


%
%


\subsection{Identical oscillators with 2- and 4-body interactions}
\label{secoa2}

Comparing Eq.~(\ref{vfield}) with Eq.~(\ref{kuracomplete3d}) we see that the equations are in agreement if we set $\vec{q} = K_1 \vec{r} + K_3 r^2 \vec{r}$ and $K_2=0$. This takes care of the inclusion of four-body interactions. However, the three-body interaction does not fit into this format
because $\vec{R}_2$ cannot be written only in terms of the order parameter $\vec{r}$. We will deal with this problem in Section III.C.

In order to avoid dealing with the general integral and differential equations for the order parameter, we simplify our analysis by restricting the model to identical particles. In this case, we can choose, without loss of generality, $G(\vec\omega) = \delta(\vec{\omega}-\vec\omega_0)= \delta(\omega_x) \delta(\omega_y) \delta(\omega_z+\omega_0)$ to obtain $\vec{r} = \vec\rho(\vec{\omega_0})$ and
\begin{equation}
	\begin{split}
		\dot{\vec{r}} = \omega_0 \, \vec r \times \hat{z}  + \frac{1}{2} (1+r^2) \vec{q}  -  (\vec{r} \cdot \vec{q} ) \vec{r}.
	\end{split}
	\label{eqm1f2}
\end{equation}   
Substituting $\vec{q} = K_1 \vec r + K_3 r^2 \vec r$ we obtain
\begin{equation}
	\begin{split}
		\dot{\vec{r}} = \omega_0 \, \vec r \times \hat{z}  + \frac{1}{2} (1-r^2)(K_1+r^2 K_3)\vec{r}.
	\end{split}
	\label{eqm1f3}
\end{equation}   

\subsection{Adding 3-body interactions}
\label{secoa3}

Three-body interactions are harder to incorporate into the reduced equations because their functional form involves the
correlation matrix $\mathbf{U}=\frac{1}{N} \sum_j ( \vec{\sigma}_j \vec{\sigma}_j^T)$ . However, as discussed in Section \ref{secprlimb}, when $r \approx 0$, its effect is to make $K_1 \rightarrow K_1 - K_2/3$, while for $r\approx 1$ it results in $K_1 \rightarrow K_1 + K_2$. 

Using these results, we propose the ansatz 
\be\label{ouransatz} K_1\to K_1-K_2/3+(4K_2/3+K_3)r^2,\ee 
which matches the limits $r\rightarrow 0$ and $r\rightarrow 1$, while interpolating between the two in a simple way. Of course, there are many other possibilities that correctly interpolate between $r=0$ and $r=1$. We tried several possibilities to include $K_2$ in the equations of motion, but this was by far the best in terms of similarity to the numerical results. This ansatz should be very accurate for $r$ close to zero or one, but slightly worse for intermediate values of $r$, as we shall see.

Under this assumption, the complete equation becomes
\begin{equation}
\dot{\vec{r}} = \omega_0 \, \vec r \times \hat{z}  + \frac{1}{2} (1-r^2) g(r) \vec{r}.
	\label{eqm1f3}
\end{equation}   
where 
\be 
g(r) = K_1 - K_2/3 + (4K_2/3 + K_3) r^2. 
\label{gr}
\ee

\section{External force}

In this Section we consider the presence of an external time-dependent force acting on the oscillators. This problem was first considered for the original Kuramoto model \cite{Sakaguchi1988,Childs2008} and later applied to networks of oscillators \cite{Hindes2015,moreira2019global,moreira2019modular}. For clarity of presentation, we set $K_2=K_3=0$ and put the higher-order terms back later. 

First, let us go back to the traditional model. Starting from 
\be 
\dot{\theta}_i = \omega_i + \frac{K_1}{N} \sum_{j=1}^N \sin{(\theta_j-\theta_i)}-F_0\sin(\theta_i-\omega_{\rm ext} t), 
\label{kuramfor}
\ee
we see that the explicit time dependence of the force can be eliminated by defining new variables $\theta_i'=\theta_i - \omega_{\rm ext} t$. The interaction terms depend on phase differences and are not affected by the transformation. We get, after dropping the primes, 
\begin{equation} 
\dot{\theta_i} = (\omega_i -\omega_{\rm ext})+ \dfrac{K_1}{N} \sum_{j=1}^{N} \sin(\theta_j - \theta_i) - F_0 \sin\theta_i.
\end{equation}

In the vector formulation, Eq.~(\ref{kuramfor}) takes the form of
\begin{equation}
	\frac{d \vec\sigma_i}{d t} =  \mathbf{W}_i \vec\sigma_{i}  +   \{(K_1\vec{r}+\vec{F}) - [\vec{\sigma}_i \cdot (K_1\vec{r}+\vec{F})] \vec\sigma_i \},
	\label{veck4a}
\end{equation}
where $\vec{F} = F_0 (\cos\omega_{\rm ext} t, \sin \omega_{\rm ext} t)$ and $\mathbf{W}_i$ is given by Eq.~(\ref{wmat}). Now we eliminate the time dependence by defining $\vec{\sigma}_i' = \mathbf{M} \vec{\sigma}_i$ where
\begin{equation}
	\mathbf{M} = \left(
	\begin{array}{cc}
		\cos(\omega_{\rm ext} t) & \sin(\omega_{\rm ext} t) \\
		-\sin(\omega_{\rm ext} t) & \cos(\omega_{\rm ext} t)
	\end{array}
	\right)
\end{equation}
is a rotation matrix. We get 
\begin{equation}
	\dot{\vec{\sigma}}_i'= \dot{\mathbf{M}} \vec{\sigma}_i + \mathbf{M}  	\dot{\vec{\sigma}}_i = \dot{\mathbf{M}} \mathbf{M}^T \mathbf{M} \vec{\sigma}_i + \mathbf{M}  	\dot{\vec{\sigma}}_i .
\end{equation}
It is easy to check that
\begin{equation}
	 \dot{\mathbf{M}} \mathbf{M}^T = 
	 \left(
	 \begin{array}{cc}
	 	0 & \omega_{\rm ext}  \\
	 	-\omega_{\rm ext}  & 0
	 \end{array}
	 \right)
\end{equation}
and that $\mathbf{M} \vec{F} = (F_0,0) \equiv \vec{F}'$. The equation for $\vec{\sigma}_i' $ becomes
\begin{equation}
	\frac{d \vec\sigma_i'}{d t} =  \mathbf{W'}_i \vec\sigma_{i}'  +   [(K_1\vec{r}+\vec{F}') - (\vec{\sigma}_i' \cdot (K_1\vec{r}+\vec{F}')) \vec\sigma_i' ],
	\label{veck4}
\end{equation}
where
\begin{equation}
	\mathbf{W'}_i = \left( 
	\begin{array}{cc}
		0 & -(\omega_i-\omega_{\rm ext}) \\
		(\omega_i-\omega_{\rm ext}) & 0
	\end{array}
	\right).
	\label{wmatt}
\end{equation}

In 3 dimensions the force can have components in all three directions and it is not clear that the time dependence can be removed by a rotation even if $\vec{F}$ is periodic. In this case, we generally have to integrate the non-autonomous equation (\ref{veck4a}) directly. In order to keep the calculations simple and derive analytical results, we choose $	\vec{F} = F_0 ( \cos(\omega_{\rm ext} t), \sin( \omega_{\rm ext} t), 0)$. In this case, we set
\begin{equation}
	\mathbf{M} = \left(
	\begin{array}{ccc}
		\cos(\omega_{\rm ext} t) & \sin(\omega_{\rm ext} t) & 0\\
		-\sin(\omega_{\rm ext} t) & \cos(\omega_{\rm ext} t) & 0\\
		0 & 0 & 1
	\end{array}
	\right)
    \label{rot}
\end{equation}
so that
\begin{equation}
	\dot{\mathbf{M}} \mathbf{M}^T = 
	\left(
	\begin{array}{ccc}
		0 & \omega_{\rm ext}  & 0\\
		-\omega_{\rm ext}  & 0 & 0\\
		0 & 0 & 0
	\end{array}
	\right).
\end{equation}
This changes $\mathbf{W}_i$ to
\begin{equation}
	\mathbf{W}_i \rightarrow \mathbf{M}^T \mathbf{W}_i \mathbf{M} +  \left( 
	\begin{array}{ccc}
		0 & \omega_{\rm ext}  &  0 \\
		-\omega_{\rm ext} & 0  &  0 \\
		0 & 0 & 0 
	\end{array}
	\right).
	\label{wmat3}
\end{equation}
For our choice of identical oscillators, $\mathbf{W}_i$ commutes to $\mathbf{M}$, since $\vec{\omega}_i=-\omega_0 \hat{z}$. In this case, we get
\begin{equation}
	\mathbf{W}_i \rightarrow  \left( 
	\begin{array}{ccc}
		0 & -(\omega_0-\omega_{\rm ext})  &  0\\
		\omega_0-\omega_{\rm ext} & 0  &  0 \\
		0 & 0 & 0 
	\end{array}
	\right)
	\label{wmat3a}
\end{equation}
and
\begin{equation}
	\vec{F}  \rightarrow \vec{F}'= \mathbf{M} \vec{F} = 
		\left(
	\begin{array}{c}
		F_0 \\
		0 \\
		0 
	\end{array}
	\right) .
\end{equation}
In short, the natural rotation in the laboratory frame is $\vec{\omega}=\omega_0\hat{z}$, where the external force is $\vec{F}=F_0(\cos(\omega_{\rm ext} t),\sin(\omega_{\rm ext} t), 0)$. We then move to a reference frame that corotates with the external force, in which the oscillators have frequency
\be 
\vec{\Omega}=(\omega_0-\omega_{\rm ext})\hat{z},
\ee 
and the external force is $\vec{F}'=(F_0,0, 0)$.

According to Eq.~(\ref{veck4}), to obtain the reduced equation for the order parameter in the presence of the external force in the rotating frame, we change $K_1 \vec{r} \rightarrow K_1 \vec{r} + \vec{F}'$ and restore the higher-order interactions. The final equation of motion becomes
\be\label{motion} \dot{\vec{r}}=\vec{\Omega}\times\vec{r}+\frac{1}{2}g(r)(1-r^2)\vec{r}+\frac{1}{2}(1+r^2)F_0 \hat{x} -F_0 x\vec{r},
\ee
where $g(r)$ is given by Eq.~(\ref{gr}).

\section{Similarity with 2D case and special solutions}

\subsection{Regrouping of coupling constants}

From Eq.~(\ref{motion}), and using $r\,dr/dt =(x\dot{x}+y\dot{y}+z\dot{z})$, it is easy to see that 
\be \frac{dr}{dt}=\left[\left(K_1-\frac{K_2}{3}\right)+\left(\frac{4K_2}{3}+K_3\right)r^2\right](1-r^2)r-xF_0\left(\frac{1}{r}-r\right).
\ee
We note that $r=1$ is always a fixed point. This is to be expected since our oscillators are identical. 

We also note that, when $z=0$, this equation is functionally identical to the one that arises in the $2$D system, a generalization 
of Eq.~(\ref{rdot2D}) in the presence of external forces, which is \cite{costa2024bifurcations}
\be 
\frac{dr}{dt}=\left[K_1+\left(K_2+K_3\right)r^2\right](1-r^2)r-xF_0\left(\frac{1}{r}-r\right).
\label{r2d}
\ee
The difference is that we have $K_1-K_2/3$ in place of $K_1$ and $4K_2/3+K_3$ in place of $K_2+K_3$. Interestingly,
3-body and 4-body interactions act additively in $2$D, as Eq.~(\ref{r2d}) depends only on $K_2+K_3$. However, in $3$D,
this symmetry is broken and $K_2$ hinders the synchronization when $r$ is close to zero and only facilitates it when $r$ 
is close to 1 \cite{fariello2024third}.

Therefore, and rather curiously, it turns out that the difference between the $2$D model and the $3$D  model in the $z=0$ plane corresponds to a regrouping of the coupling constants. Notice that $\left(K_1-\frac{K_2}{3}\right)+\left(\frac{4K_2}{3}+K_3\right)=K_1+K_2+K_3$, so the regrouping preserves their sum.

\subsection{Fixed points}

In cartesian coordinates, Eq.~(\ref{motion}) reads as:
\be \dot{x}=\Omega y+\frac{1}{2}g(r)(1-r^2)x+\frac{1}{2}(1+r^2-2x^2)F_0, \label{eqxx}\ee
\be \dot{y}=-\Omega x+\frac{1}{2}g(r)(1-r^2)y-xyF_0,\label{eqyy}\ee
\be \dot{z}=\frac{1}{2}g(r)(1-r^2)z-xzF_0 \label{eqzz}\ee
where $g(r)$ is given by Eq.~(\ref{gr}).

Some equilibrium points are available. For example, if $F_0>\Omega$ we have two:
\be
\label{trivial1} 
x_\pm=\pm\frac{1}{F_0}\sqrt{F_0^2-\Omega^2},\quad y=-\frac{\Omega}{F_0},\quad z=0.
\ee
The corresponding Jacobian eigenvalues are: $-\sqrt{F_0^2-\Omega^2}$ (twice) and $-\sqrt{F_0^2-\Omega^2}-K_1-K_2-K_3$ for $x_+$; $\sqrt{F_0^2-\Omega^2}$ (twice) and $\sqrt{F_0^2-\Omega^2}-K_1-K_2-K_3$ for $x_-$. When the coupling constants are positive, one of these fixed points is stable, and the other is unstable. At the line $F=\Omega$ they collide in a saddle-node bifurcation.

On the other hand, if $F_0<\Omega$ we again have two fixed points (not in the $z=0$ plane this time),
\be\label{trivial2} x=0, \quad y=-\frac{F_0}{\Omega}, \quad z=\pm\frac{1}{\Omega}\sqrt{\Omega^2-F_0^2},\ee
both of them with the same Jacobian eigenvalues, two imaginary ones, $\pm i\sqrt{\Omega^2-F_0^2}$, and a real one, $-(K_1+K_2+K_3)$. The eigenvector corresponding to the real eigenvalue is in the $yz$ plane, so the stability is determined by the sign of $K_1+K_2+K_3$.

\subsection{Total coherence and periodic solutions}

If $r^2=x^2+y^2+z^2=1$, then the equations of motion simplify to
\be \dot{x}=\Omega y+(1-x^2)F_0,\ee
\be \dot{y}=-\Omega x-xyF_0,\ee
\be \dot{z}=-xzF_0.\ee

\begin{figure}
    \centering
    \includegraphics[width=0.4\linewidth, height=0.415\linewidth]{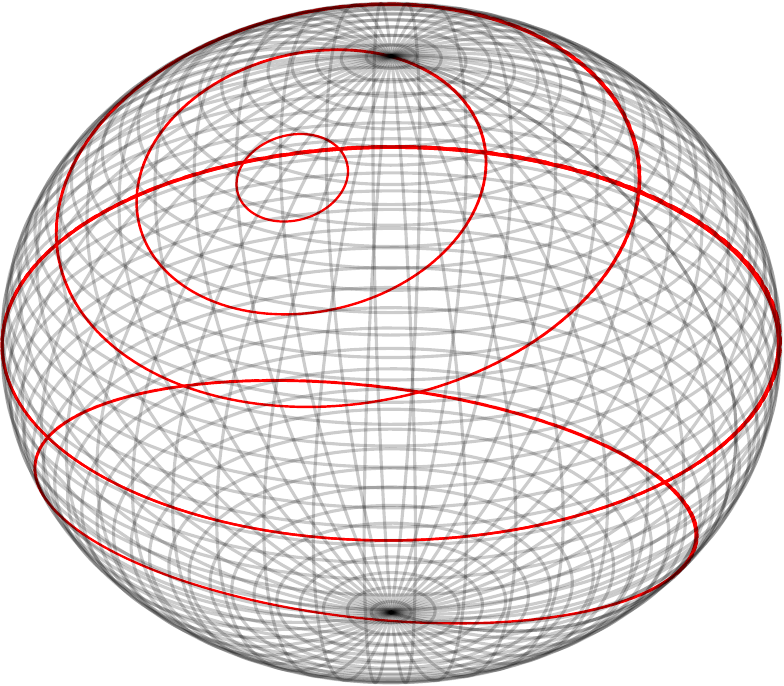}
    \caption{Examples of perfectly coherent orbits for different initial conditions with parameters $K_1 = 1$, $K_2 = 7$, $K_3 = 0$, $\Omega = 2$ and $F = 1$.}
    \label{fig:trajetorias}
\end{figure}

The general solution can be explicitly found as 
\be x(t)=\frac{\alpha}{F_0}\frac{A\cos(\alpha t)-\sin(\alpha t)}{A\sin(\alpha t)+\cos(\alpha t)+B},\ee
where $A, B$ are arbitrary constants and
$\alpha=\sqrt{\Omega^2-F_0^2}.$ (Notice that $A=i$, $B=0$ recovers 
Eq.~(\ref{trivial1}), and $B\to\infty$ recovers Eq.~(\ref{trivial2}).) The other components are given in terms of $x(t)$ by 
\be y(t)=\frac{F_0}{\Omega}(x^2-1)+\frac{1}{\Omega}\dot{x}\ee
and 
\be z(t)=z(0)e^{-F_0\int_0^t x(t')dt'},\ee
with the constant $z(0)$ being related to previous parameters as 
\be 
z(0)=\frac{\alpha}{F_0\Omega}\sqrt{B^2F_0^2-\Omega^2(A^2+1)},
\ee
in order to guarantee $r=1$. Fig.~\ref{fig:trajetorias} shows examples of trajectories in this family. Notice that
they all run on the surface of the sphere.

\section{Bifurcations}

Eqs.~(\ref{eqxx})-(\ref{eqzz}) always admit the solution $z=0$. As we have seen, in this plane, the dynamics is similar to the $2$D model. Correspondingly, the present model displays bifurcation scenarios that are similar to the ones uncovered in \cite{costa2024bifurcations} (except for the fact that in that work the oscillators were not identical, as they are here, but had different natural frequencies).

\subsection{Hopf bifurcation}

In a supercritical Hopf bifurcation, a small periodic orbit is born out of a stable fixed point that becomes unstable. However, the Hopf bifurcation we found in $3$D is subcritical, where an unstable periodic orbit collides with a stable fixed point, making it unstable. We find the locus of this bifurcation by restricting the dynamics to the $z=0$ plane and computing the corresponding Jacobian as 
\begin{equation}\label{jacobian}
	J_{xy} = \begin{pmatrix}
		\dfrac{\partial \dot{x}}{\partial x} & \dfrac{\partial \dot{x}}{\partial y} \\
		\dfrac{\partial \dot{y}}{\partial x} & \dfrac{\partial \dot{y}}{\partial y}
	\end{pmatrix}.
\end{equation}

In a 2D Hopf bifurcation, the Jacobian eigenvalues are of the form $\pm i\lambda$, so its trace is zero and its determinant is positive. Solving the system of equations $\dot{x}=\dot{y}={\rm Tr}(J_{xy})=0$ for the three unknowns $x,y,F$, we find at once the values of the fixed point $(x_{\rm Hopf},y_{\rm Hopf})$ and of $F_{\rm Hopf}$ as functions of $\Omega$. However, this must be done numerically (see next section). 

\subsection{Saddle-node bifurcation}

Here, it is easier to work with polar coordinates in the $x$-$y$ plane. Using
$ \dot{r} = \dot{x} \cos\theta + \dot{y} \sin\theta $ and
$ r\dot{\theta} = \dot{y} \cos\theta - \dot{x} \sin\theta$, we get
\be \dot{r} = \frac{1}{2} (1-r^2) \left[ r g(r) + F_0 \cos\theta \right] \ee
and
\be r \dot{\theta} = -\Omega r - \frac{1}{2} (1+r^2) F_0 \sin\theta. \ee

When $r=1$ the equation for $\theta$ becomes $\dot{\theta} = -\Omega  - F_0 \sin\theta$, leading to a SNIPER bifurcation at $F_0=\Omega$. When $r\neq 1$, the fixed points are solutions of the system
\be r g(r) = -F_0 \cos\theta \ee
and
\be \frac{2 \Omega r}{1+r^2} = -F_0 \sin\theta. \ee
Squaring these equations and summing the results we get rid of $\theta$ and find the relation $a(r)=b(r)$ 
where $a(r)= \frac{F_0^2}{r^2}$ and $b(r) = g(r)^2 + \frac{4 \Omega^2}{(1+r^2)^2}.$ There may be multiple solutions to the above equation. When both sides have the same slope,
\be a'(r)=b'(r),\ee
two equilibrium points coalesce in a saddle-node bifurcation. By solving these two equations, we can numerically find $r_{\rm SN}$ and $F_{\rm SN}$ as functions of $\Omega$. 

\subsection{Takens-Bogdanov point}

At the point where the Hopf curve meets the saddle-node curve, both eigenvalues of $J_{xy}$ are zero. This is called the Takens-Bogdanov (TB) point. Solving the equations $\dot{x}=\dot{y}=\det(J_{xy})={\rm Tr}(J_{xy})=0$ we can obtain the coordinates $(F_{\rm TB},\Omega_{\rm TB})$ of this point in the $F\times \Omega$ plane.

For $K_1=1$, $K_2=7$, $K_3=0$, this is approximately expressed as $F_{\rm TB}\approx 0.27$ and $\Omega_{\rm TB}\approx 0.40$. For $K_3=0$, we plot $F_{\rm TB}$ as a function of $K_1$ for some values of $K_2$ below. We do not show the corresponding plot for $\Omega_{\rm TB}$, because it happens that $\Omega_{\rm TB} \simeq F_{\rm TB}$ at the TB point. The plot suggests that the TB point disappears for $K_1>K_2/3$ and that it runs off to infinity as $K_1\to -K_2$.

\begin{figure}
    \centering
    \includegraphics[width=0.5\linewidth]{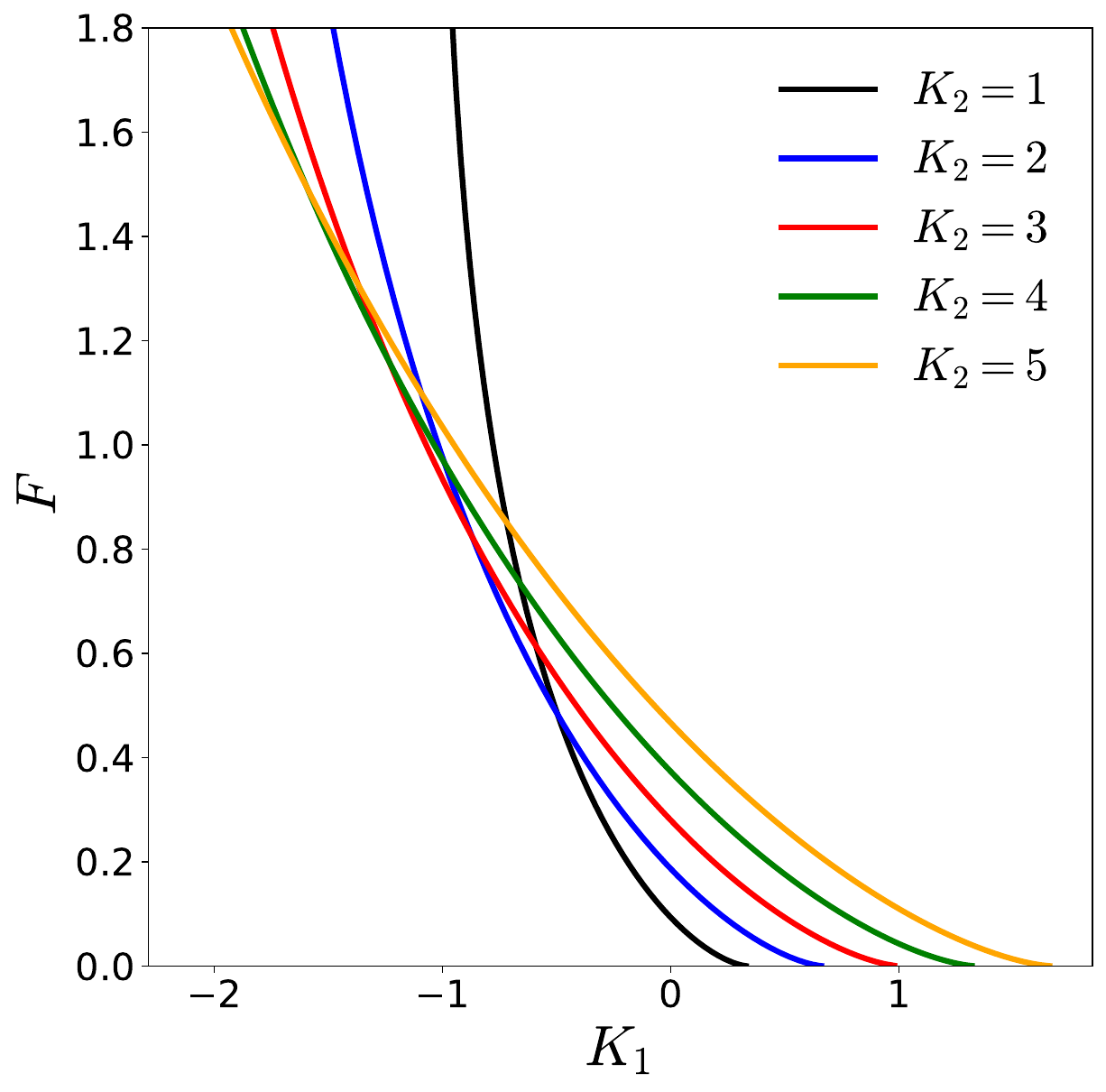}
    \caption{$F_{\rm TB}$ as a function of $K_1$ for $K_3=0$ at the Takens-Bogdanov point for several values of $K_2$. The curves terminate at $K_1=K_2/3$.}
    \label{fig:takens}
\end{figure}

\begin{figure}
\centering
\includegraphics[width=0.4\linewidth]{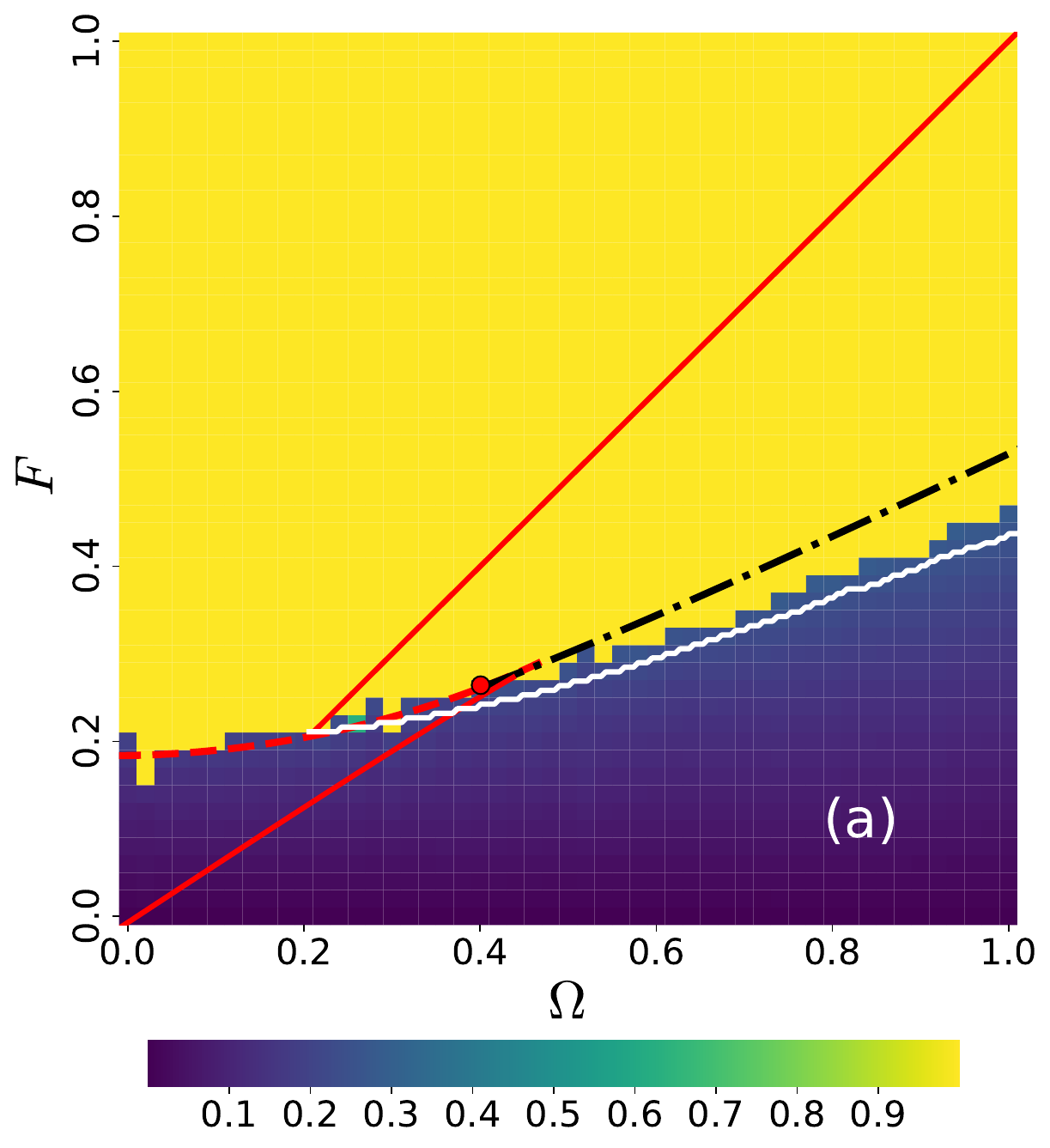}
\includegraphics[width=0.4\linewidth]{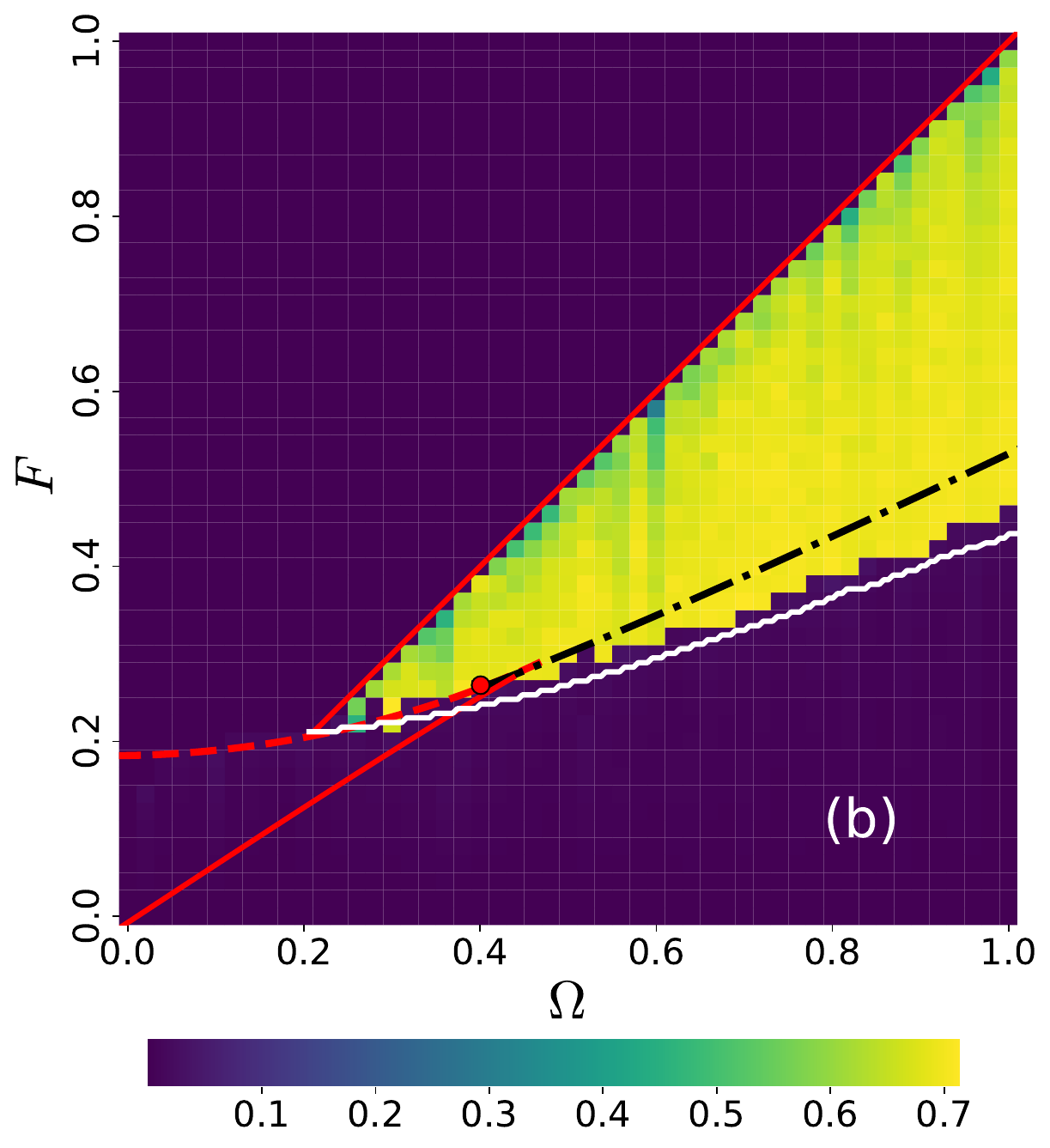}
\caption{Heatmap of (a) time-averaged order parameter $\langle r \rangle $ and (b) time-averaged largest deviation of $\vec{r}$ components for a system of $N = 2000$ oscillators with coupling parameters $K_1 = 1$, $K_2 =7$ and $K_3 = 0$. Overlapping lines are bifurcation manifolds found via ansatz. Full lines are SNIPER bifurcations. The dashed one is a common saddle-node and the dot-dashed is a subcritical Hopf. The white line is the numerical solution of the reduced equations with ansatz $f_P$ (see text).}
\label{fig:heatmatpK1K7}
\end{figure}

\begin{figure}
\centering
\includegraphics[width=0.9\linewidth]{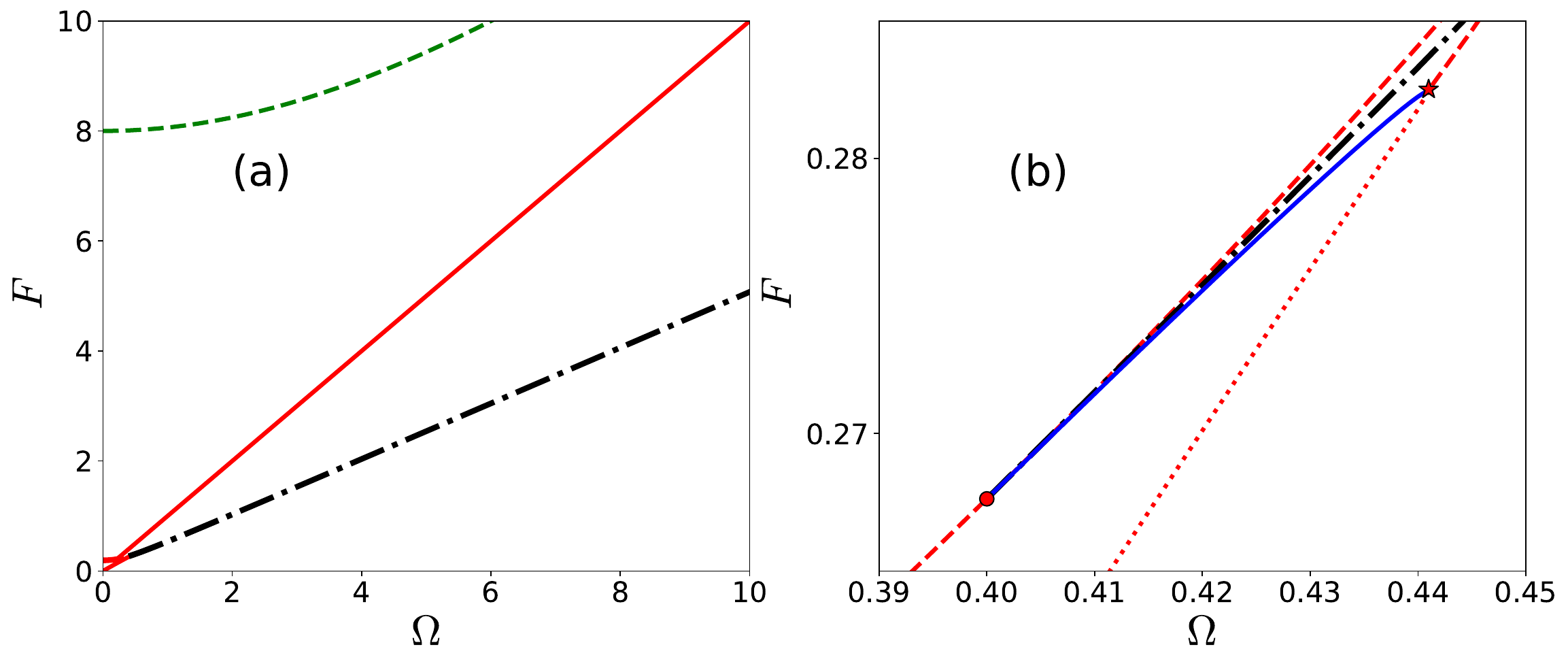}
\caption{Bifurcation diagram in the $F \times \Omega$ plane for $K_1=1$, $K_2=7$ and $K_3=0$. (a) For large values of
$F$ a second SN bifurcation manifold can be observed (green dashed curve), similar to the 2D case. The SNIPER curve is
shown by the continuous red curve and the subcritical Hopf bifurcation by the black dot-dashed line. (b) Zoom of panel (a) for
small values of $F$ and $\Omega$, demonstrating the SN curve displayed in Fig.~\ref{fig:heatmatpK1K7} (red dashed line) and the line of homoclinic bifurcations emerging from the SN at the TB point (filled circle) and ending at the Bautin point (filled star). }
\label{fig:bifuk1k7}
\end{figure}

\begin{figure}
\centering
\includegraphics[width=0.5\linewidth]{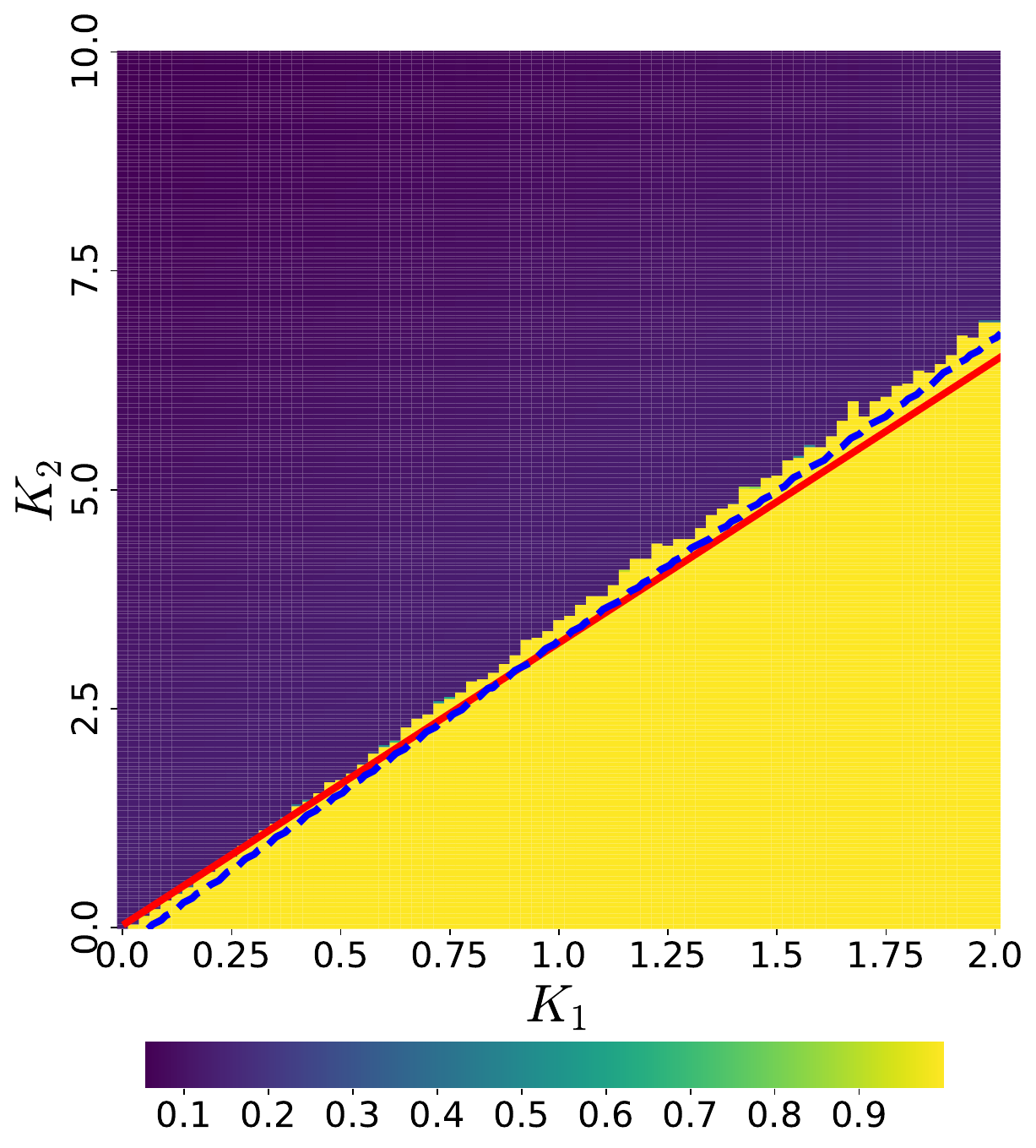}
\caption{Heatmap of time-averaged order parameter $\langle r \rangle $ for a system of $N = 5000$ oscillators with coupling parameters $F = 0.2$, $\Omega =1$ and $K_3 = 0$. Continuous line is the Hopf bifurcation obtained via reduced equations with $f_Y$ and dashed line is the numerical solution via $f_P$. }
\label{fig:heatmatpK1K2}
\end{figure}

\section{Simulations}

In this section, we illustrate our theoretical findings comparing direct simulations of the Kuramoto model and the reduced equations derived with the help of the Ott-Antonsen ansatz and the approximation introduced in Section \ref{secoa3}. Since the role of four-body interaction is less interesting, in the sense that it amounts to change $K_1$ to $K_1+r^2 K_3$ and always facilitates synchronization, we set $K_3=0$ and focus on the roles of $K_2$ and the external force.

Fig.~\ref{fig:heatmatpK1K7}(a) shows a heatmap of the order parameter in the $F \times \Omega$ plane for $K_1=1$, $K_2=7$ and $K_3=0$ obtained by direct integration of the Kuramoto equations (\ref{motion}). Continuous and dashed lines show the bifurcation manifolds computed analytically (see the figure caption for details). Fig.~\ref{fig:heatmatpK1K7}(b) shows the largest mean square deviation (msd) of the components of $\vec{r}$, indicating whether $\vec{r}$ is a fixed point (small msd) or a periodic orbit (large msd). Simulations for the Kuramoto model were performed with 2000 oscillators using a variable time step Runge-Kutta fourth-order algorithm. The agreement between theory and simulation is very good, except close to the subcritical Hopf bifurcation. We hypothesize that this mismatch is partly due to the nature of the bifurcation, as the basin of attraction of the stable fixed point shrinks and becomes hard to find close to the critical point. However, the mismatch might also be due in part to our approximation, Eq.~(\ref{ouransatz}), for including three-body terms, and the choice of ansatz $f_Y$. In order to check the results of reduced equations via $f_P$, we numerically integrated Eq.~(\ref{eqm1fa}) with $\vec{q} = g(r) \vec{r} + F \hat{x}$, $\vec{\omega}= \Omega \hat{z}$ and $\vec{r} = h(\rho) \vec{\rho}$. The result is the white line, which indeed provides a better approximation to the simulations.

Fig.~\ref{fig:bifuk1k7} shows the bifurcation diagram in the $F \times \Omega$ plane. Panel (a) shows the second branch of the SN bifurcations that exist for large values of $F$. Panel (b) is a zoom of the diagram for small values of $F$ and $\Omega$, showing in detail all the bifurcation manifolds found in the system, including a homoclinic one that is found by dedicated numerical routines (this is a short manifold that starts at the Taken-Bogdanov point and ends at the so-called Bautin point).

Fig.~\ref{fig:heatmatpK1K2} shows the heatmap of the order parameter in the $K_1 \times K_2$ plane for $F=0.2$ and $\Omega=1$, computed 
from Eq.~(\ref{motion}). The heatmap has a point in common with
Fig.~\ref{fig:bifuk1k7} at $K_1=1$, $K_2=7$. The red continuous line shows the theoretical locus of the Hopf bifurcation. Once again a small mismatch between theory and simulation is observed near the critical line. This time, the numerical solution obtained via $f_P$ (dashed line) is just as good as that provided by $f_Y$. We must, however, note that the border line coming from the simulation is not very clear cut. 

\section{Conclusions and Discussion}

The extension of the Kuramoto model to more dimensions has become a topic of intense research in the past years \cite{lohe2009non,dai2021d,fariello2024third,Fariello2024a,de2023generalized,2019continuous}. One of the most interesting cases is that of the sphere, where the transition to synchronization is discontinuous \cite{2019continuous} and phases can be interpreted as orientation angles of a flying bird or drone \cite{chandra2020extensions}.  

The problem is amenable to an exact theoretical description via dimensional reduction when all oscillators are identical and only two-body interactions are present. However, the presence of inhomogeneous units, external forces, and higher-order interactions poses a challenge. In this work, we suggest a way of including three- and four-body interactions into the equations of motion for the case of identical oscillators. This allows us to obtain exact results for the locus of various bifurcations that organize the dynamics of the system. Our suggestion is validated by agreement with numerical simulations. 

We find that the plane $z=0$ is invariant under the dynamics and that on this plane the equations of motion are structurally identical to the ones from the usual Kuramoto model, up to a renormalization of the coupling constants in which we have $K_1-K_2/3$ instead of $K_1$ and $4K_2/3+K_3$ instead of $K_2+K_3$. So, the role of $K_2$ in the spherical model is, on the one hand, to shift the synchronization transition to higher values when $K_1$ is small and to facilitate it when $K_1$ is large \cite{fariello2024third}. We also found a two-parameter family of periodic solutions outside the $z=0$ plane, representing real 3D oscillations.

An interesting detail of this analysis is the choice of dimensional reduction used to derive our equations. The dimensional reduction on the sphere is more complicated than its analogue for the usual Kuramoto model. The goal is to replace the dynamics of a large number of oscillators with that of the order parameter $\vec{r}$ and an auxiliary variable $\vec{\rho}$, the so-called ansatz parameter. Two such methods have been put forward as below.

The first, by Chandra et al \cite{chandra2019complexity}, assumes that the density of oscillators with natural frequency $\vec{\omega}$ has the same functional form as that of the Ott-Antonsen ansatz for the Kuramoto model \cite{Ott2008} and adjusts the exponents of the terms in order to satisfy the continuity equation. The result, denoted as $f_P(\vec\omega,\hat{r},t)$ as specified in Eq.~(\ref{ansatz1}), is proportional to the Poisson kernel \cite{lipton2021Kuramoto} and $\vec{\rho}$ should satisfy Eq.~(\ref{eqm1fa}). The relation between $\vec{r}$ and $\vec{\rho}$ is a weighted average over the distribution of natural frequencies, Eq.~(\ref{r1}).

The second method consists of expressing the density as a series in spherical harmonics and making an ansatz for the coefficients \cite{barioni2021ott}, similar to the derivation of the Ott-Antonsen ansatz \cite{Ott2008}. The result is represented by $f_Y(\vec\omega,\hat{r},t)$ as indicated in Eq.~(\ref{anssol}) and the ansatz parameter $\vec{\rho}$ should satisfy the same Eq.~(\ref{eqm1fa}), as long as the interaction vector $\vec{q}$ is displaced as $\vec{q} \rightarrow \vec{q} + (K/4)[(\hat{\rho}-\hat{\sigma})\cdot \vec{r}] \hat{\sigma}$ \cite{barioni2021ott}. Although such a displacement does not change the velocity field, it is unusual. This inconvenience is addressed by requiring that $|\vec{\rho}(0)|=1$. With this choice, the density reduces to $\delta(\hat{\sigma}-\hat{\rho})$ and the displacement disappears. The advantage of this method is that $\vec{r}$ and $\vec{\rho}$ are related by a simple average over the distribution of natural frequencies, Eq.~(\ref{r2}).

Numerical simulations show that, for strictly identical oscillators, the performance of $f_P$ is slightly better than that of $f_Y$, however, this difference disappears as soon as some heterogeneity is introduced in the natural frequencies. For this reason, our analytical treatment uses $f_Y$ for its simplicity.

Future extensions of this work might consider nonidentical oscillators with different frequency distributions and the introduction of frustration, similar to the Kuramoto-Sakaguchi model \cite{sakaguchi1986soluble}. This can be done by changing the coupling constants to rotation matrices, as proposed in \cite{buzanello2022matrix,de2023generalized}. In the sphere, such matrices would generally not commute with the frequency matrices $\mathbf{W}_i$ and can lead to interesting dynamical properties.

\begin{acknowledgments}
This work was partly supported by FAPESP, grants 2021/14335-0 (MAMA), 2023/03917-4 (GSC) and 2023/15644-2 (MN), and also by CNPq, grants 303814/2023-3 (MAMA) and 304986/2022-4 (MN). We thank the Coaraci Supercomputer for computer time (FAPESP grant 2019/17874-0) and the Center for Computing in Engineering and Sciences at Unicamp (FAPESP grant 2013/08293-7).
\end{acknowledgments}


\clearpage
\enlargethispage{1\baselineskip}


\newpage
\begin{thebibliography}{10}
	
	\bibitem{Moiseff181}
	A.~Moiseff and J.~Copeland, ``Firefly synchrony: A behavioral strategy to
	minimize visual clutter,'' {\em Science}, vol.~329, no.~5988, pp.~181--181,
	2010.
	
	\bibitem{hadar1985head}
	U.~Hadar, T.~J. Steiner, and F.~Clifford~Rose, ``Head movement during listening
	turns in conversation,'' {\em Journal of Nonverbal Behavior}, vol.~9,
	pp.~214--228, 1985.
	
	\bibitem{neda2000sound}
	Z.~N{\'e}da, E.~Ravasz, Y.~Brechet, T.~Vicsek, and A.-L. Barab{\'a}si, ``The
	sound of many hands clapping,'' {\em Nature}, vol.~403, no.~6772,
	pp.~849--850, 2000.
	
	\bibitem{Mackay1997}
	W.~A. Mackay, ``{Synchronized neuronal oscillations and their role in motor
		processes.},'' {\em Trends in cognitive sciences}, vol.~1, no.~13646613,
	pp.~176--183, 1997.
	
	\bibitem{cumin2007generalising}
	D.~Cumin and C.~Unsworth, ``Generalising the kuramoto model for the study of
	neuronal synchronisation in the brain,'' {\em Physica D: Nonlinear
		Phenomena}, vol.~226, no.~2, pp.~181--196, 2007.
	
	\bibitem{Deco2011}
	G.~Deco, A.~Buehlmann, T.~Masquelier, and E.~Hugues, ``{The Role of Rhythmic
		Neural Synchronization in Rest and Task Conditions},'' {\em Frontiers in
		Human Neuroscience}, vol.~5, p.~4, feb 2011.
	
	\bibitem{Schmidt2015}
	R.~Schmidt, K.~J.~R. LaFleur, M.~A. de~Reus, L.~H. van~den Berg, and M.~P.
	van~den Heuvel, ``{Kuramoto model simulation of neural hubs and dynamic
		synchrony in the human cerebral connectome.},'' {\em BMC neuroscience},
	vol.~16, no.~1, p.~54, 2015.
	
	\bibitem{Michaels704}
	D.~C. Michaels, E.~P. Matyas, and J.~Jalife, ``Mechanisms of sinoatrial
	pacemaker synchronization: a new hypothesis.,'' {\em Circulation Research},
	vol.~61, no.~5, pp.~704--714, 1987.
	
	\bibitem{oliveira2014huygens}
	H.~M. Oliveira and L.~V. Melo, ``Huygens synchronization of two pendulum
	clocks,'' {\em arXiv preprint arXiv:1410.7926}, 2014.
	
	\bibitem{Pantaleone2002}
	J.~Pantaleone, ``Synchronization of metronomes,'' {\em American Journal of
		Physics}, vol.~70, no.~10, pp.~992--1000, 2002.
	
	\bibitem{winfree1967biological}
	A.~T. Winfree, ``Biological rhythms and the behavior of populations of coupled
	oscillators,'' {\em Journal of theoretical biology}, vol.~16, no.~1,
	pp.~15--42, 1967.
	
	\bibitem{Kuramoto1975}
	Y.~Kuramoto, ``{Self-entrainment of a population of coupled non-linear
		oscillators},'' in {\em International Symposium on Mathematical Problems in
		Theoretical Physics}, pp.~420--422, Berlin/Heidelberg: Springer-Verlag, 1975.
	
	\bibitem{Kuramoto1984}
	Y.~Kuramoto, ``{Chemical Waves},'' in {\em Chemical Oscillations, Waves, and
		Turbulence}, pp.~89--110, Springer Berlin Heidelberg, 1984.
	
	\bibitem{harding1994photosensitive}
	G.~F. Harding and P.~M. Jeavons, {\em Photosensitive epilepsy}.
	\newblock No.~133, Cambridge University Press, 1994.
	
	\bibitem{Reece2012}
	J.~B. Reece, {\em {Campbell biology : concepts {\&} connections}}.
	\newblock San Francisco, CA.: Benjamin Cummings, 2012.
	
	\bibitem{Sakaguchi1988}
	H.~Sakaguchi, ``Cooperative phenomena in coupled oscillator systems under
	external fields,'' {\em Progress of Theoretical Physics}, vol.~79, no.~1,
	pp.~39--46, 1988.
	
	\bibitem{Childs2008}
	L.~M. Childs and S.~H. Strogatz, ``{Stability diagram for the forced Kuramoto
		model},'' {\em Chaos}, vol.~18, no.~4, pp.~1--9, 2008.
	
	\bibitem{Hindes2015}
	J.~Hindes and C.~R. Myers, ``Driven synchronization in random networks of
	oscillators,'' {\em Chaos: An Interdisciplinary Journal of Nonlinear
		Science}, vol.~25, no.~7, p.~073119, 2015.
	
	\bibitem{Lizarraga2020}
	J.~U.~F. Lizarraga and M.~A.~M. de~Aguiar, ``Synchronization and spatial
	patterns in forced swarmalators,'' {\em Chaos (Woodbury, N.Y.)}, vol.~30,
	p.~053112, May 2020.
	
	\bibitem{PhysRevLett.131.030401}
	T.~Murtadho, S.~Vinjanampathy, and J.~Thingna, ``Cooperation and competition in
	synchronous open quantum systems,'' {\em Phys. Rev. Lett.}, vol.~131,
	p.~030401, Jul 2023.
	
	\bibitem{odor2023synchronization}
	G.~{\'O}dor, I.~Papp, S.~Deng, and J.~Kelling, ``Synchronization transitions on
	connectome graphs with external force,'' {\em Frontiers in Physics}, vol.~11,
	p.~1150246, 2023.
	
	\bibitem{Ott2008}
	E.~Ott and T.~M. Antonsen, ``{Low dimensional behavior of large systems of
		globally coupled oscillators},'' {\em Chaos}, vol.~18, no.~3, pp.~1--6, 2008.
	
	\bibitem{Rodrigues2016}
	F.~A. Rodrigues, T.~K. D.~M. Peron, P.~Ji, and J.~Kurths, ``{The Kuramoto model
		in complex networks},'' {\em Physics Reports}, vol.~610, pp.~1--98, 2016.
	
	\bibitem{lohe2009non}
	M.~Lohe, ``Non-abelian {K}uramoto models and synchronization,'' {\em Journal of
		Physics A: Mathematical and Theoretical}, vol.~42, no.~39, p.~395101, 2009.
	
	\bibitem{2019continuous}
	S.~Chandra, M.~Girvan, and E.~Ott, ``Continuous versus discontinuous
	transitions in the d-dimensional generalized kuramoto model: Odd d is
	different,'' {\em Physical Review X}, vol.~9, no.~1, p.~011002, 2019.
	
	\bibitem{Tanaka2014}
	T.~Tanaka, ``Solvable model of the collective motion of heterogeneous particles
	interacting on a sphere,'' {\em New Journal of Physics}, vol.~16, 01 2014.
	
	\bibitem{chandra2019complexity}
	S.~Chandra, M.~Girvan, and E.~Ott, ``Complexity reduction ansatz for systems of
	interacting orientable agents: Beyond the kuramoto model,'' {\em Chaos: An
		Interdisciplinary Journal of Nonlinear Science}, vol.~29, no.~5, p.~053107,
	2019.
	
	\bibitem{lipton2021Kuramoto}
	M.~Lipton, R.~Mirollo, and S.~H. Strogatz, ``The kuramoto model on a sphere:
	Explaining its low-dimensional dynamics with group theory and hyperbolic
	geometry,'' {\em Chaos: An Interdisciplinary Journal of Nonlinear Science},
	vol.~31, no.~9, p.~093113, 2021.
	
	\bibitem{barioni2021complexity}
	A.~E.~D. Barioni and M.~A. de~Aguiar, ``Complexity reduction in the 3d kuramoto
	model,'' {\em Chaos, Solitons \& Fractals}, vol.~149, p.~111090, 2021.
	
	\bibitem{barioni2021ott}
	A.~E.~D. Barioni and M.~A. de~Aguiar, ``Ott--antonsen ansatz for the
	d-dimensional kuramoto model: A constructive approach,'' {\em Chaos: An
		Interdisciplinary Journal of Nonlinear Science}, vol.~31, no.~11, p.~113141,
	2021.
	
	\bibitem{tanaka2011multistable}
	T.~Tanaka and T.~Aoyagi, ``Multistable attractors in a network of phase
	oscillators with three-body interactions,'' {\em Physical Review Letters},
	vol.~106, no.~22, p.~224101, 2011.
	
	\bibitem{bick2016chaos}
	C.~Bick, P.~Ashwin, and A.~Rodrigues, ``Chaos in generically coupled phase
	oscillator networks with nonpairwise interactions,'' {\em Chaos: An
		Interdisciplinary Journal of Nonlinear Science}, vol.~26, no.~9, 2016.
	
	\bibitem{battiston2020networks}
	F.~Battiston, G.~Cencetti, I.~Iacopini, V.~Latora, M.~Lucas, A.~Patania, J.-G.
	Young, and G.~Petri, ``Networks beyond pairwise interactions: Structure and
	dynamics,'' {\em Physics Reports}, vol.~874, pp.~1--92, 2020.
	
	\bibitem{dutta2023impact}
	S.~Dutta, A.~Mondal, P.~Kundu, P.~Khanra, P.~Pal, and C.~Hens, ``Impact of
	phase lag on synchronization in frustrated kuramoto model with higher-order
	interactions,'' {\em Physical Review E}, vol.~108, no.~3, p.~034208, 2023.
	
	\bibitem{leon2024higher}
	I.~Le{\'o}n, R.~Muolo, S.~Hata, and H.~Nakao, ``Higher-order interactions
	induce anomalous transitions to synchrony,'' {\em Chaos: An Interdisciplinary
		Journal of Nonlinear Science}, vol.~34, no.~1, 2024.
	
	\bibitem{ganmor2011sparse}
	E.~Ganmor, R.~Segev, and E.~Schneidman, ``Sparse low-order interaction network
	underlies a highly correlated and learnable neural population code,'' {\em
		Proceedings of the National Academy of sciences}, vol.~108, no.~23,
	pp.~9679--9684, 2011.
	
	\bibitem{petri2014homological}
	G.~Petri, P.~Expert, F.~Turkheimer, R.~Carhart-Harris, D.~Nutt, P.~J. Hellyer,
	and F.~Vaccarino, ``Homological scaffolds of brain functional networks,''
	{\em Journal of The Royal Society Interface}, vol.~11, no.~101, p.~20140873,
	2014.
	
	\bibitem{giusti2015clique}
	C.~Giusti, E.~Pastalkova, C.~Curto, and V.~Itskov, ``Clique topology reveals
	intrinsic geometric structure in neural correlations,'' {\em Proceedings of
		the National Academy of Sciences}, vol.~112, no.~44, pp.~13455--13460, 2015.
	
	\bibitem{reimann2017cliques}
	M.~W. Reimann, M.~Nolte, M.~Scolamiero, K.~Turner, R.~Perin, G.~Chindemi,
	P.~D{\l}otko, R.~Levi, K.~Hess, and H.~Markram, ``Cliques of neurons bound
	into cavities provide a missing link between structure and function,'' {\em
		Frontiers in computational neuroscience}, vol.~11, p.~266051, 2017.
	
	\bibitem{sizemore2018cliques}
	A.~E. Sizemore, C.~Giusti, A.~Kahn, J.~M. Vettel, R.~F. Betzel, and D.~S.
	Bassett, ``Cliques and cavities in the human connectome,'' {\em Journal of
		computational neuroscience}, vol.~44, pp.~115--145, 2018.
	
	\bibitem{grilli2017higher}
	J.~Grilli, G.~Barab{\'a}s, M.~J. Michalska-Smith, and S.~Allesina,
	``Higher-order interactions stabilize dynamics in competitive network
	models,'' {\em Nature}, vol.~548, no.~7666, pp.~210--213, 2017.
	
	\bibitem{ghosh2024chimeric}
	R.~Ghosh, U.~K. Verma, S.~Jalan, and M.~D. Shrimali, ``Chimeric states induced
	by higher-order interactions in coupled prey--predator systems,'' {\em Chaos:
		An Interdisciplinary Journal of Nonlinear Science}, vol.~34, no.~6, 2024.
	
	\bibitem{sanchez2019high}
	A.~Sanchez-Gorostiaga, D.~Baji{\'c}, M.~L. Osborne, J.~F. Poyatos, and
	A.~Sanchez, ``High-order interactions distort the functional landscape of
	microbial consortia,'' {\em PLoS Biology}, vol.~17, no.~12, p.~e3000550,
	2019.
	
	\bibitem{benson2016higher}
	A.~R. Benson, D.~F. Gleich, and J.~Leskovec, ``Higher-order organization of
	complex networks,'' {\em Science}, vol.~353, no.~6295, pp.~163--166, 2016.
	
	\bibitem{de2020social}
	G.~F. de~Arruda, G.~Petri, and Y.~Moreno, ``Social contagion models on
	hypergraphs,'' {\em Physical Review Research}, vol.~2, no.~2, p.~023032,
	2020.
	
	\bibitem{fariello2024third}
	R.~Fariello and M.~A. {de Aguiar}, ``Third order interactions shift the
	critical coupling in multidimensional kuramoto models,'' {\em Chaos, Solitons
		\& Fractals}, vol.~187, p.~115467, 2024.
	
	\bibitem{skardal2020higher}
	P.~S. Skardal and A.~Arenas, ``Higher order interactions in complex networks of
	phase oscillators promote abrupt synchronization switching,'' {\em
		Communications Physics}, vol.~3, no.~1, p.~218, 2020.
	
	\bibitem{Strogatz2019higher}
	M.~Lipton, R.~Mirollo, and S.~H. Strogatz, ``On higher dimensional generalized
	kuramoto oscillator systems,'' {\em arXiv preprint arXiv:1907.07150}, 2019.
	
	\bibitem{buzanello2022matrix}
	G.~L. Buzanello, A.~E.~D. Barioni, and M.~A. de~Aguiar, ``Matrix coupling and
	generalized frustration in kuramoto oscillators,'' {\em Chaos: An
		Interdisciplinary Journal of Nonlinear Science}, vol.~32, no.~9, p.~093130,
	2022.
	
	\bibitem{de2023generalized}
	M.~A.~M. de~Aguiar, ``Generalized frustration in the multidimensional kuramoto
	model,'' {\em Phys. Rev. E}, vol.~107, p.~044205, Apr 2023.
	
	\bibitem{moreira2019global}
	C.~A. Moreira and M.~A. de~Aguiar, ``Global synchronization of partially forced
	kuramoto oscillators on networks,'' {\em Physica A: Statistical Mechanics and
		its Applications}, vol.~514, pp.~487--496, 2019.
	
	\bibitem{moreira2019modular}
	C.~A. Moreira and M.~A. de~Aguiar, ``Modular structure in c. elegans neural
	network and its response to external localized stimuli,'' {\em Physica A:
		Statistical Mechanics and its Applications}, vol.~533, p.~122051, 2019.
	
	\bibitem{costa2024bifurcations}
	G.~S. Costa, M.~Novaes, and M.~A. de~Aguiar, ``Bifurcations in the kuramoto
	model with external forcing and higher-order interactions,'' {\em Chaos: An
		Interdisciplinary Journal of Nonlinear Science}, vol.~34, no.~12, 2024.
	
	\bibitem{dai2021d}
	X.~Dai, K.~Kovalenko, M.~Molodyk, Z.~Wang, X.~Li, D.~Musatov, A.~Raigorodskii,
	K.~Alfaro-Bittner, G.~Cooper, G.~Bianconi, {\em et~al.}, ``D-dimensional
	oscillators in simplicial structures: odd and even dimensions display
	different synchronization scenarios,'' {\em Chaos, Solitons \& Fractals},
	vol.~146, p.~110888, 2021.
	
	\bibitem{Fariello2024a}
	R.~Fariello and M.~A. de~Aguiar, ``Exploring the phase diagrams of
	multidimensional kuramoto models,'' {\em Chaos, Solitons \& Fractals},
	vol.~179, p.~114431, 2024.
	
	\bibitem{chandra2020extensions}
	S.~Chandra, {\em Extensions of the Kuramoto model: from spiking neurons to
		swarming drones}.
	\newblock PhD thesis, University of Maryland, College Park, 2020.
	
	\bibitem{sakaguchi1986soluble}
	H.~Sakaguchi and Y.~Kuramoto, ``A soluble active rotater model showing phase
	transitions via mutual entertainment,'' {\em Progress of Theoretical
		Physics}, vol.~76, no.~3, pp.~576--581, 1986.
	
\end{thebibliography}
\end{document}